\documentclass[aps,prb,twocolumn,showpacs,superscriptaddress]{revtex4}
\usepackage{amsmath}
\usepackage{amssymb}
\usepackage{epsfig}
\usepackage{color}
\usepackage{graphicx,amsmath}

\begin{document}
\title{Common quantum phase transition in quasicrystals and heavy-fermion metals}
\author{V. R. Shaginyan}\email{vrshag@thd.pnpi.spb.ru}
\affiliation{Petersburg Nuclear Physics Institute, Gatchina,
188300, Russia}\affiliation{Clark Atlanta University, Atlanta, GA
30314, USA}\author{A. Z. Msezane} \affiliation{Clark Atlanta
University, Atlanta, GA 30314, USA} \author{K. G.
Popov}\affiliation{Komi Science Center, Ural Division, RAS,
Syktyvkar, 167982, Russia} \author{G. S.
Japaridze}\affiliation{Clark Atlanta University, Atlanta, GA 30314,
USA}\author{V. A. Khodel} \affiliation{Russian Research Centre
Kurchatov Institute, Moscow, 123182, Russia} \affiliation{McDonnell
Center for the Space Sciences \& Department of Physics, Washington
University, St.~Louis, MO 63130, USA}

\begin{abstract}
Extraordinary new materials named quasicrystals and characterized
by noncrystallographic rotational symmetry and quasiperiodic
translational properties have attracted scrutiny. Study of
quasicrystals may shed light on the most basic notions related to
the quantum critical state observed in heavy-fermion metals. We
show that the electronic system of some quasicrystals is located at
the fermion condensation quantum phase transition without tuning.
In that case the quasicrystals possess the quantum critical state
with the non-Fermi liquid behavior which in magnetic fields
transforms into the Landau Fermi-liquid one. Remarkably, the
quantum critical state is robust despite the strong disorder
experienced by the electrons. We also demonstrate for the first
time that quasicrystals exhibit the typical scaling behavior of
their thermodynamic properties such as the magnetic susceptibility,
and belong to the famous family of heavy-fermion metals. Our
calculated thermodynamic properties are in good agreement with
recent experimental observations.
\end{abstract}

\pacs{ 71.23.Ft, 71.27.+a, 05.30.Rt} \maketitle

When encountering exciting behavior of strongly correlated metals,
we anticipate to learn more about quantum critical physics. Such an
opportunity may be provided by quasicrystals (QCs) \cite{sch}.
These, characterized by the absence of translational symmetry in
combination with good atomic arrangement and rotational symmetry,
can be viewed as materials located between crystalline and
disordered solids. QCs, approximants and related complex metallic
phases reveal very unusual mechanical, magnetic, electronic
transport and thermodynamic properties. The aperiodicity of QCs
plays an important role at the formation of the properties since
the band electronic structure governed by the Bloch theorem cannot
be well defined. As an example, QCs exhibit a high resistivity
although the density of states (DOS) at the Fermi energy is not
small \cite{trans}. One expects transport properties to be defined
by a small diffusivity of electrons which occupy a new class of
states denoted as "critical states", neither being extended nor
localized, and making the velocity of charge carriers very low
\cite{trans}. Associated with these critical states, characterized
by an extremely degenerate confined wave function, are the
so-called "spiky" DOS\, \cite{fuj,fuj1}. These predicted DOS are
corroborated by experiments revealing that single spectra of the
local DOS demonstrate a spiky DOS\, \cite{wid,rwid}. Clearly these
spiky states are associated with flat bands \cite{flat,guy}. On one
hand, we expect the properties related to the itinerate states
governed by the spiky states of QCs to coincide with that of
heavy-fermion metals, while on the other hand, the pseudo localized
states may result in those of amorphous materials. Therefore, the
question of how quasicrystalline order influences the electronic
properties in quasicrystals, whether these resemble those of
heavy-fermion (HF) metals or those of amorphous materials, is of
crucial importance.

Recently, experimental measurements on the gold-aluminium-ytterbium
quasicrystal $\rm Au_{51}Al_{34}Yb_{15}$ with a six-dimensional
lattice parameter $a_d=0.7448$ nm have revealed a quantum critical
behavior with the unusual exponent $\alpha\simeq0.51$ defining the
divergency of the magnetic susceptibility $\chi\propto T^{-\alpha}$
as temperature $T\to0$\, \cite{QCM}. The measurements have also
exposed that the observed non-Fermi liquid (NFL) behavior
transforms into Landau Fermi liquid (LFL) under the application of
a tiny magnetic field $H$, while it exhibits the robustness against
hydrostatic pressure; the quasicrystal shows also metallic behavior
with the $T-$dependent part $\Delta \rho$ of the resistivity,
$\Delta \rho\propto T$, at low temperatures \cite{QCM}. All these
facts challenge theory to explain a quantum criticality of the
gold-aluminum-ytterbium QC characterized by the unusual exponent
and robust against hydrostatic pressure but destroyed by tiny
magnetic fields.

In this communication we uncover that a quantum critical point of
$\rm Au_{51}Al_{34}Yb_{15}$, generating the NFL behavior, is a
fermion condensation quantum phase transition (FCQPT)
\cite{pr,shag} and also present the first explanation of the low
temperature thermodynamics in magnetic fields. We explain the
robustness of the quantum critical behavior against the hydrostatic
pressure, and how the application of a weak magnetic field destroys
the behavior and makes the system transit from the NFL to LFL
behavior. We also demonstrate that there is a general mechanism
underlying the NFL behavior of HF metals and quasicrystals, leading
to a scaling behavior.

We start with constructing a model to explain the challenging
behavior of  the gold-aluminum-ytterbium QC. Taking into account
that the spiky states are associated to flat bands \cite{flat}
which are the generic signature of FCQPT, we safely assume that the
electronic system of the gold-aluminum-ytterbium QC $\rm
Au_{51}Al_{34}Yb_{15}$ is located very near FCQPT \cite{pr}. Thus,
$\rm Au_{51}Al_{34}Yb_{15}$ turns out to be located at FCQPT
without tuning this substance with the pressure, magnetic field
etc. We expect that the system exhibits the robustness of its
critical behavior against the hydrostatic pressure since the
hydrostatic pressure does not change the topological structure of
QC leading to the spiky DOS and, correspondingly, flat bands. As we
will see, the spiky DOS cannot prevent the field-induced Fermi
liquid state.

To study the low temperature thermodynamic and scaling behavior, we
use the model of homogeneous heavy-fermion liquid \cite{pr}. This
model avoids the complications associated with the anisotropy of
solids and considering both the thermodynamic properties and NFL
behavior by calculating the effective mass $M^*(T,H)$ as a function
of temperature $T$ and magnetic field $H$. To study the behavior of
the effective mass $M^*(T,H)$, we use the Landau equation for the
quasiparticle effective mass. The only modification is that in our
formalism the effective mass is no longer constant but depends on
temperature and magnetic field. For the model of homogeneous HF
liquid at finite temperatures and magnetic fields, this equation
takes the form \cite{pr,shag,mig100,land}
\begin{eqnarray}
\nonumber \frac{1}{M^*_{\sigma}(T,
H)}&=&\frac{1}{M}+\sum_{\sigma_1}\int\frac{{\bf p}_F{\bf
p}}{p_F^3}F_
{\sigma,\sigma_1}({\bf p_F},{\bf p}) \\
&\times&\frac{\partial n_{\sigma_1} ({\bf
p},T,H)}{\partial{p}}\frac{d{\bf p}}{(2\pi)^3}, \label{HC1}
\end{eqnarray}
where $M$ is a bare electron mass, $F_{\sigma,\sigma_1}({\bf
p_F},{\bf p})$ is the Landau interaction, which depends on Fermi
momentum $p_F$, momentum $p$ and spin $\sigma$. Here we use the
units where $\hbar=k_B=1$. The Landau interaction has the form
\cite{land}
\begin{equation}\label{lampl}
F_{\sigma,\sigma'}({\bf p},{\bf p'})= \frac{\delta^2 E[n]}{\delta
n_{\sigma}({\bf p})\delta n_{\sigma'}({\bf p'})},
\end{equation}
where $E[n]$ is the system energy, which is a functional of the
quasiparticle distribution function $n$\,\, \cite{pr,mig100,land}.
It can be expressed as
\begin{equation}
n_{\sigma}({\bf p},T)=\left\{ 1+\exp \left[\frac{(\varepsilon({\bf
p},T)-\mu_{\sigma})}T\right]\right\} ^{-1},\label{HC2}
\end{equation}
where $\varepsilon({\bf p},T)$ is the single-particle spectrum. In
our case, the chemical potential $\mu$ depends on the spin due to
Zeeman splitting $\mu_{\sigma}=\mu\pm \mu_BH$, $\mu_B$ is Bohr
magneton. The single-particle spectrum is a variational derivative
of the system energy $E[n_{\sigma}({\bf p})]$ with respect to the
quasiparticle distribution function or occupation numbers $n$,
\begin{equation}
\varepsilon({\bf p})=\frac{\delta E[n({\bf p})]}{\delta
n}.\label{EN}
\end{equation}
In our case $F$ is fixed by the condition that the system is
situated at FCQPT. The variational procedure, being applied to the
functional $E[n_{\sigma}({\bf p},T)]$, gives using the form for
$\varepsilon({\bf p},T)=\varepsilon_\sigma({\bf p},T)\equiv
\varepsilon[n_{\sigma}({\bf p},T)]$,
\begin{equation}\label{epta}
\frac{\partial\varepsilon_\sigma({\bf p},T)}{\partial{\bf
p}}=\frac{p}{M}+\sum_{\sigma_1}\int F_ {\sigma,\sigma_1}({\bf
p},{\bf p}_1)\frac{\partial n_{\sigma_1}({\bf p}_1,T)}{\partial
{\bf p}}\frac{d^3p_1}{(2\pi)^3}.
\end{equation}
Equations \eqref{HC2} and \eqref{epta} constitute the closed set
for self-consistent determination of $\varepsilon_\sigma({\bf
p},T)$ and $n_{\sigma}({\bf p},T)$. The sole role of the Landau
interaction is to bring the system to FCQPT point, where
$M^*\to\infty$ at $T=0$ and $H=0$, and the Fermi surface alters its
topology so that the effective mass acquires temperature and field
dependence \cite{pr,shag,mig100,ckz}. Provided that the Landau
interaction is an analytical function, at the Fermi surface the
momentum-dependent part of the Landau interaction can be taken in
the form of truncated power series $F=aq^2+ bq^3+cq^4+...$, where
${\bf q}={\bf p}_1-{\bf p}_2$, $a,b$ and $c$ are fitting parameters
which are defined by the condition that the system is at FCQPT.
Close to the Fermi momentum $p_F$, the electron spectrum
$\varepsilon(p)$, given by Eq. \eqref{epta} with the above
interaction $F$, behaves as
$\varepsilon(p)-\mu\propto(p-p_F)^3$\,\, \cite{ckz,pr}. A direct
inspection of Eq. \eqref{HC1} shows that at $T=0$ and $H=0$, the
sum of the first term and the second one on the right side
vanishes, since $1/M^*(T\to0)\to 0$ provided that the system is
located at FCQPT \cite{pr,ckz}. In case of analytic Landau
interaction at finite $T$ the right hand side is proportional
$F^{\prime}(M^*)^2T^2$, where $F^{\prime}$ is the first derivative
of $F$ with respect to $q$ at $q\to0$. Calculations of the
corresponding integrals can be found in textbooks, see e.g.
\cite{lanl2} Thus, we have $1/M^*\propto (M^*)^2T^2$, and obtain
\cite{pr,ckz}
\begin{equation}
M^*(T)\simeq a_TT^{-2/3}.\label{MT}
\end{equation}
At finite $T$, the application of magnetic field $H$ drives system
to the LFL region with
\begin{equation}
M^*(H)\simeq a_HH^{-2/3}.\label{MB}
\end{equation}
On the other hand, an analytic interaction $F$ can lead to the
general topological form of the spectrum
$\varepsilon(p)-\mu\propto(p-p_b)^2(p-p_F)$ with $(p_b<p_F)$ and
$(p_F-p_b)/p_F\ll1$, that makes $M^*\propto T^{-1/2}$, and creates
a quantum critical point \cite{sqrt}. As we shall see below, the
same critical point is generated by the interaction $F(q)$
represented by an integrable over $x$ nonanalytic function with
$q=\sqrt{p_1^2+p_2^2-2xp_1p_2}$ and $F(q\to0)\to \infty$\,
\cite{pr,pr1}. The both cases lead to $M^*\propto T^{-1/2}$, and
Eq. \eqref{MT} becomes
\begin{equation}
M^*(T)\simeq a_T T^{-1/2}.\label{MTT}
\end{equation}
In the same way, we obtain
\begin{equation}
M^*(H)\simeq a_H H^{-1/2},\label{MBB}
\end{equation}
with $a_T$ and $a_H$ are parameters. Taking into account that Eq.
\eqref{MTT} leads to the spiky DOS with the vanishing of spiky
structure with increasing temperature \cite{dkss}, as it is
observed in quasicrystals \cite{rwid,QCM}, we assume that the
general form of $\varepsilon(p)$ produces the behavior of $M^*$,
given by Eqs. \eqref{MTT} and \eqref{MBB}, and is realized in
quasicrystals which can be viewed as a generalized form of common
crystals \cite{ron}. We note that the behavior
$1/M^*\propto\chi^{-1}\propto T^{1/2}$ is in good agreement with
$\chi^{-1}\propto T^{0.51}$ observed experimentally \cite{QCM}. Our
explanation is consistent with the robustness of the exponent
$0.51$ against the hydrostatic pressure \cite{QCM} since the
robustness is guaranteed by the unique singular DOS of QCs that
survives under the application of pressure
\cite{fuj,fuj1,rwid,flat,QCM}. Then, the nonanalytic Landau
interaction can also serve as the good approximation, generating
the observed behavior of the effective mass. We speculate that the
nonanalytic interaction is generated by the nonconservation of the
quasimomentum in QCs, making the Landau interaction $F(q)$ a
nonlocal function of momentum $q$. Such a function can be
approximated by a nonanalytic one.

A few remarks on the transport properties of QC are in order here.
In calculations of low-temperature resistivity, we employ a
two-band model, one of which is occupied by heavy quasiparticles,
with the effective mass given by Eq. \eqref{MTT}, while the second
band possesses a LFL quasiparticles with a $T-$independent
effective mass \cite{prb}. As a result, we find that  the
quasiparticles width $\gamma\propto T$ and that the $T-$dependent
part of the resistivity $\Delta\rho\propto T$. This observation is
in accordance with experimental facts\cite{QCM}.

\begin{figure}[!ht]
\begin{center}
\includegraphics [width=0.50\textwidth]{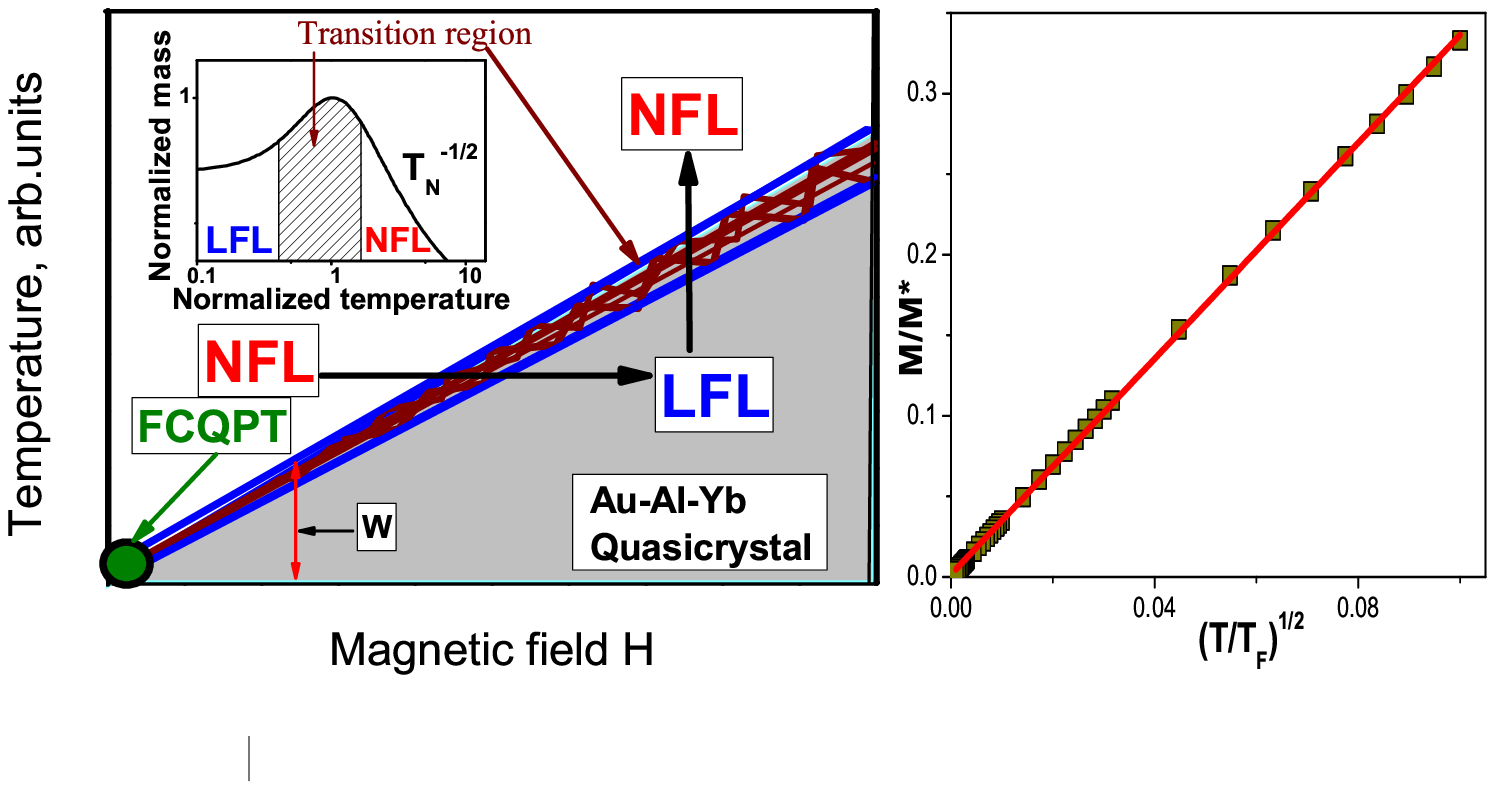}
\vspace*{-1.5cm}
\end{center}
\caption{(color online). Left panel. $T-H$ phase diagram of $\rm
Au_{51}Al_{34}Yb_{15}$ versus magnetic field $H$ as the control
parameter. The vertical and horizontal arrows show LFL-NFL and
NFL-LFL transitions at fixed $H$ and $T$ respectively. At $H=0$ and
$T=0$ the system is at FCQPT shown by the solid circle. The common
width of the LFL and the transition regions $W\propto T$ are shown
by the double arrows. Inset shows a schematic plot of the
normalized effective mass versus the normalized temperature.
Transition region, where $M^*_N$ reaches its maximum at $T/T_{\rm
max}=1$, is shown by the hatched area. The right panel reports the
dimensionless inverse effective mass $M/M^*$ versus dimensionless
temperature $(T/T_F)^{1/2}$. The line is a linear fit.}\label{fig1}
\end{figure}

At finite $H$ and $T$ near FCQPT, the solutions of Eq. \eqref{HC1}
$M^*(T,H)$ can be well approximated by a simple universal
interpolating function. A deeper insight into the behavior of
$M^*(T,H)$ can be achieved using some "internal" scales. Namely,
near FCQPT the solutions of Eq. \eqref{HC1} exhibit a universal
scaling behavior so that $M^*(T,H)$ reaches it maximum value
$M^*_M$ at some temperature $T_{\rm max}\propto H$ \cite{pr,ckz}.
It is convenient to introduce the internal scales $M^*_M$ and
$T_{\rm max}$ to measure the effective mass and temperature,
respectively. Thus, we divide the effective mass $M^*$ and the
temperature $T$ by their maximal values, $M^*_M$ and $T_{\rm max}$
respectively. This generates the normalized effective mass
$M^*_N=M^*/M^*_M$ and temperature $T_N=T/T_{\rm max}$ \cite{pr}.
Near FCQPT the normalized solution of Eq. \eqref{HC1} $M^*_N(T_N)$
with a nonanalytic Landau interaction can be well approximated by a
simple universal interpolating function. The interpolation occurs
between the LFL ($M^*\propto a+ bT^2$) and NFL ($M^*\propto
T^{-1/2}$) regimes and represents the universal scaling behavior of
$M^*_N(T_N)$
\begin{equation}M^*_N(T_N)\approx c_0\frac{1+c_1T_N^2}{1+c_2T_N^{5/2}}.
\label{UN2}
\end{equation}
Here $a$ and $b$ are constants, $c_0=(1+c_2)/(1+c_1)$, $c_1$ and
$c_2$ are fitting parameters. The inset to the left panel of Fig.
\ref{fig1} shows the scaling behavior of the normalized effective
mass. It is seen from the inset, that the common width $W$ of the
LFL and the transition region $W\propto T$ vanish as $H\to0$ since
$T_{\rm max}\propto H$. In the same way, the common width of the
NFL and the transition region tends to zero as soon as $T\to0$.

Now we construct the schematic phase diagram of the
gold-aluminum-ytterbium QC $\rm Au_{51}Al_{34}Yb_{15}$. The phase
diagram is reported in Fig. \ref{fig1}, left panel. The magnetic
field $H$ plays the role of the control parameter, driving the
system outwards FCQPT that occurs at $H=0$ and $T=0$ without tuning
since the QC critical state is formed by singular density of states
\cite{fuj,fuj1,rwid,flat,QCM}. It follows from Eq. \eqref{UN2} and
seen from the left panel of Fig. \ref{fig1}, that at fixed
temperatures the increase of $H$ drives the system along the
horizontal arrow from NFL state to LFL one. On the contrary, at
fixed magnetic field and increasing temperatures the system
transits along the vertical arrow from LFL state to NFL one. The
inset to the left panel demonstrates the behavior of the normalized
effective mass $M^*_N$ versus normalized temperature $T_N$
following from Eq. \eqref{UN2}. The $T^{-1/2}$ regime is marked as
NFL since contrary to the LFL case, where the effective mass is
constant, the effective mass depends strongly on temperature. It is
seen that the temperature region $T_N\sim 1$ signifies a transition
regime between the LFL behavior with almost constant effective mass
and the NFL one, given by $T^{-1/2}$ dependence. Thus, temperatures
$T\simeq T_{\rm max}$, shown by arrows in the inset and the main
panel, can be regarded as the transition regime between LFL and NFL
states. The common width $W$ of the LFL transition regions
$W\propto T$ is shown by the heavy arrow. These theoretical results
are in good agreement with the experimental facts \cite{QCM}. The
right panel of Fig. \ref{fig1} illustrates the behavior of the
dimensionless inverse effective mass $M/M^*$ versus the
dimensionless temperature $(T/T_F)^{1/2}$, where $T_F$ is the Fermi
temperature of electron gas. To calculate $M/M^*$, we use a model
Landau functional \cite{pr,pr1}
\begin{eqnarray}
\nonumber E[n(p)]&=&\int\frac{{\bf p}^2}{2M}\frac{d{\bf
p}}{(2\pi)^3}+\frac{1}
{2}\int V({\bf p}_1-{\bf p}_2)\\
&\times&n({\bf p}_1)n({\bf p}_2) \frac{d{\bf p}_1d{\bf
p}_2}{(2\pi)^6},\label{HC6}
\end{eqnarray}
with the Landau interaction
\begin{equation}
V({\bf p})=g_0\frac{\exp(-\beta_0\sqrt{{\bf
q}^2+\gamma^2})}{\sqrt{{\bf q}^2+\gamma^2}},\label{HC7}
\end{equation}
where the parameters $g_0$ and $\beta_0$ are fixed by the
requirement that the system be located at FCQPT. The interaction at
$\gamma=0$ becomes nonanalytic. It is worthy of note that the other
nonanalytic interactions lead to the same behavior of $M/M^*$, see
e.g. \cite{pr1}.

\begin{figure}[!ht]
\begin{center}
\includegraphics [width=0.40\textwidth]{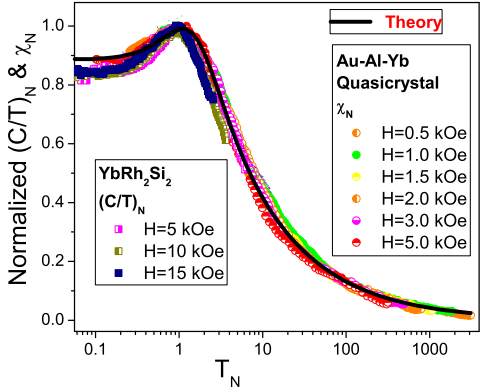}
\end{center}
\caption{(color online). The normalized specific heat $(C/T)_N$ and
normalized magnetic susceptibility $\chi_N$ extracted from
measurements in magnetic fields $H$ on $\rm YbRh_2Si_2$
\cite{pikul} and on $\rm Au_{51}Al_{34}Yb_{15}$ \cite{QCM},
respectively. The magnetic fields are given in the figure. Our
calculations are depicted by the solid curve tracing the scaling
behavior of $(C/T)_N=\chi_N=M^*_N$ given by Eq. \eqref{UN2}.}
\label{fig2}
\end{figure}
To demonstrate this, we apply Eq. \eqref{EN} to construct
$\varepsilon(p)$ using the functional \eqref{HC6}. Taking into
account that $\varepsilon(p\simeq p_F)-\mu\simeq p_F(p-p_F)$ and
integrating over the angle variables, we obtain
\begin{equation}
\frac1{M^*}=\frac1{M}+\frac{\partial}{\partial p}
\int\left[\Phi(p+p_1)-\Phi(|p-p_1|)\right]\frac{n(p_1,T)p_1dp_1}{2p_F^2\pi^2}.
\label{EMF}\end{equation} Here the derivative on the right hand
side of Eq. \eqref{EMF} is taken at $p=p_F$ and
\begin{equation}
\int^{p+p_1}_{|p-p_1|}V(z,\gamma=0)zdz=\Phi(p+p_1)-\Phi(|p-p_1|).\label{EMV}
\end{equation}
The derivative $\partial\Phi(|p-p_1|)/\partial p|_{p\to
p_F}=(p_F-p_1)/(|p_F-p_1|)\partial\Phi(z)/\partial z$ becomes a
discontinuous function at $p_1\to p_F$, provided that
$\partial\Phi(z)/\partial z$ is finite (or integrable if the
function tends to infinity) at $z\to 0$. As a result, the right
hand side of Eq. \eqref{EMF} becomes proportional $M^*T$ and
\eqref{EMF} reads $1/M^*\propto M^*T$, making $M^*\propto
T^{-1/2}$. Calculations of the corresponding integrals, entering
Eq. \eqref{EMF}, can be found in textbooks, see e.g. \cite{lanl2}.

The analytic Landau interaction \eqref{HC7} with $\gamma>0$ makes
$M/M^*\propto T^{0.5}$ at elevated temperatures, while at $T\to0$
the system demonstrates the LFL behavior \cite{sqrt,pr}. This
interaction can serve as model one to describe the behavior of the
quasicrystal's crystalline approximant $\rm
Au_{51}Al_{35}Yb_{14}$\, \cite{QCM}.
\begin{figure} [! ht]
\begin{center}
\includegraphics [width=0.40\textwidth]{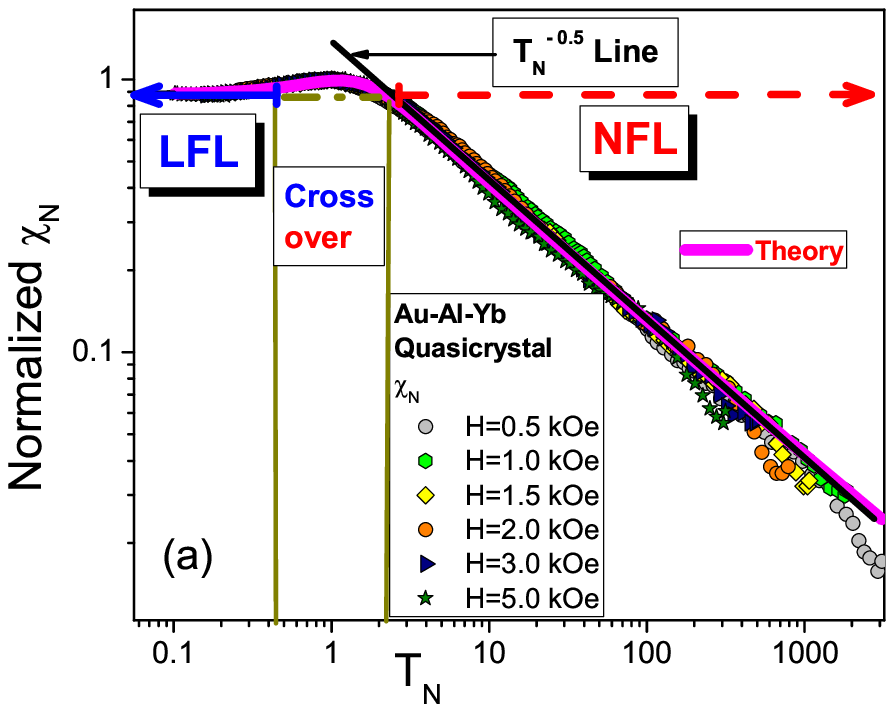}
\includegraphics [width=0.47\textwidth]{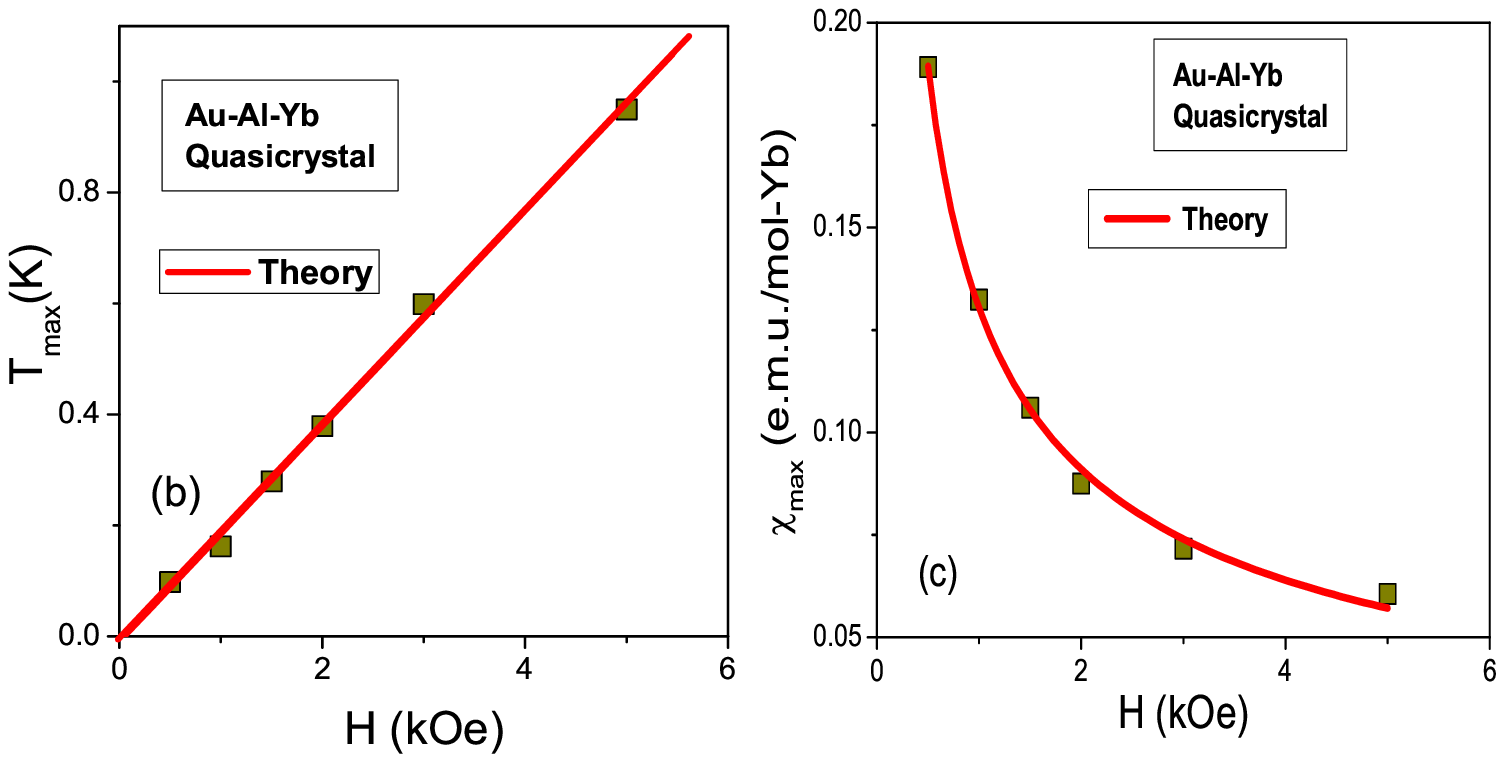}
\end{center}
\caption{(color online). (a): Temperature dependence on the double
logarithmic scale of the magnetic susceptibility $\chi_N$ at
different magnetic fields \cite{QCM} shown in the legend. The LFL
region and NFL one are shown by the solid and dashed arrows,
respectively. The solid line depicts $\chi_N\propto T_N^{-0.5}$
behavior. (b): The temperatures $T_{\rm max}$ at which the maxima
of $\chi$ (see Fig. \ref{fig1}) are located. The solid line
represents the function $T_{\rm max}=aH$, $a$ is a fitting
parameter. (c): The maxima $\chi_{\rm max}$ versus magnetic field
$H$. The solid curve is approximated by $\chi_{\rm max}=tH^{-1/2}$,
see Eq. \eqref{MBB}, $t$ is a fitting parameter. }\label{fig345}
\end{figure}
The approximant $\rm Au_{51}Al_{35}Yb_{14}$ shows the LFL behavior
at low temperatures, $\chi^{-1}\propto a+bT^{0.51}$ with the
conventional LFL behavior of the resistivity \cite{QCM}. We
interpret this behavior of $\chi^{-1}$ through the absence of the
unique electronic state of QCs, which results in the shift of the
electronic system of the approximant from FCQPT into the LFL
region. Such a behavior is achieved by making the interaction
\eqref{HC7} an analytic function with $\gamma>0$ as soon as the
quasicrystal is transformed into its crystalline approximant. The
finite $\gamma$, creating the LFL behavior at $T=0$, makes $T_{\rm
max}$ finite even at $H=0$. Then, it follows from Eq. \eqref{UN2}
that $1/M^{*}\propto\chi^{-1}\propto a+bT^{1/2}$ and the
approximant is to demonstrate the conventional LFL behavior:
$\Delta\rho\propto T^2$. The same result is acquired by
transforming the spectrum as follows,
$\varepsilon(p)-\mu\propto([p-p_b]^2+\gamma^2)(p-p_F)$
\,\cite{sqrt}.

We now investigate the behavior of $\chi$ as a function of
temperature at fixed magnetic fields. The effective mass $M^*(T,H)$
can be measured in experiments for $M^*(T,H)\propto \chi$ where
$\chi$ is the ac or dc magnetic susceptibility. If the
corresponding measurements are carried out at fixed magnetic field
$H$ then, as it follows from Eq. \eqref{UN2}, $\chi$ reaches the
maximum $\chi_{\rm max}$ at some temperature $T_{\rm max}$. Upon
normalizing both $\chi$ and the specific heat $C/T$ by their peak
values at each field $H$  and the corresponding temperatures by
$T_{\rm max}$, we observe from Eq. \eqref{UN2} that all the curves
merge into a single one, thus demonstrating a scaling behavior
typical for HF metals \cite{pr}. As seen from Fig. \ref{fig2},
$\chi_N$ extracted from measurements on $\rm Au_{51}Al_{34}Yb_{15}$
\,\cite{QCM} shows the scaling behavior given by Eq. \eqref{UN2}
and agrees well with our calculations shown by the solid curve over
four orders of magnitude in the normalized temperature.

In order to validate the phase diagram Fig. \ref{fig1}, we focus on
the LFL, NFL and the transition LFL-NFL regions exhibited by the
QC. To this end, we display in Fig. \ref{fig345} (a) the normalized
$\chi_N$ on the double logarithm scale. As seen from Fig.
\ref{fig345} (a), $\chi_N$ extracted from the measurements is not a
constant, as would be for a LFL. The two regions (the LFL region
and NFL one), separated by the transition region, as depicted by
the hatched area in the inset of Fig. \ref{fig1}, are clearly seen
in Fig. \ref{fig345} (a) illuminating good agreement between the
theory and measurements. The straight lines in Fig. \ref{fig345}
(a) outline both the LFL and NFL behaviors of $\chi_N\propto const$
and $\chi_N\propto T_N^{-1/2}$, and are in good agreement with the
behavior of $M^*_N$ displayed in the inset of Fig. \ref{fig1}. In
Fig. \ref{fig345}, (b), the solid squares denote temperatures
$T_{\rm max}(H)$ at which the maxima $\chi_{\rm max}$ of $\chi(T)$
and, (c), the corresponding values of the maxima $\chi_{\rm
max}(H)$ occur. It is seen that the agreement between the theory
and experiment is good in the entire magnetic field domain. It is
also seen from Fig. \ref{fig345} (b) that $T_{\rm max}\propto H$;
thus a tiny magnetic field $H$ destroys the NFL behavior hereby
driving the system to the LFL region. This behavior is consistent
with the phase diagram displaced in Fig. \ref{fig1}: at increasing
temperatures ($T_N\simeq 1$) the LFL state first converts into the
transition one and then disrupts into the NFL state, while at given
magnetic field $H$ the width $W\propto T$.

In summary, we have established for the first time that $\rm
Au_{51}Al_{34}Yb_{15}$ quasicrystal exhibits the typical scaling
behavior of its thermodynamic properties, and belongs to the famous
family of heavy-fermion metals. We have also demonstrated that the
quantum critical physics of the quasicrystal is universal, and
emerges regardless of the underlying microscopic details of the
quasicrystal.

This work was supported by U.S. DOE, Division of Chemical Sciences,
Office of Basic Energy Sciences, Office of Energy Research, AFOSR,
and by the projects \#12-Y-1-1010 of the Ural Division of RAS and
\#12-P-1-1014 of RAS.


\begin{thebibliography}{99}

\bibitem{sch} D. Schechtman, I. Blech, D. Gratias, and J. W. Cahn,
\prl {\bf 53}, 1951 (1984).

\bibitem{trans} D. Mayou and G. T. de Laissardi\`ere, {\it Quantum
transport in quasicrystals and complex metallic alloys}, In
Quasicrystals, series "Handbook of Metal Physics", (Eds. T.
Fujiwara, Y. Ishii, Elsevier Science, 2008), pp. 209-265.

\bibitem{fuj} T. Fujiwara and T. Yokokawa, \prl {\bf 66}, 333 (1991).

\bibitem{fuj1} T. Fujiwara, {\it in Physical Properties of Quasicrystals},
(ed. Stadnik, Z. M., Springer, 1999).

\bibitem{wid} R. Widmera, O. Gr\"oninga, P. Ruffieuxa, and P.
Gr\"oninga, Phil. Mag. {\bf 86}, 781 (2006).

\bibitem{rwid} R. Widmer, P. Gr\"oning, M. Feuerbacher, and O.
Gr\"oning, \prb {\bf 79}, 104202 (2009).

\bibitem{flat} T. Fujiwara, S. Yamamoto, and G. Trambly de
Laissardi\`ere, \prl {\bf 71}, 4166 (1993).

\bibitem{guy} G. Trambly de Laissardi\`ere,  Z. Kristallogr. {\bf
224}, 123 (2009).

\bibitem{QCM} K. Deguchi, S. Matsukawa, N. K. Sato, T. Hattori,
K. Ishida, H. Takakura, and T. Ishimasa, Nature Materials {\bf 11},
1013 (2012).

\bibitem{pr} V. R. Shaginyan, M. Ya. Amusia, A. Z. Msezane, and
K. G. Popov,  Phys. Rep. {\bf 492}, 31 (2010).

\bibitem{shag} V. R. Shaginyan, Physics of Atomic Nuclei {\bf 74},
1107 (2011).

\bibitem{mig100}  V. A. Khodel, J. W. Clark, and M. V. Zverev,
Physics of Atomic Nuclei {\bf 74}, 1237 (2011).

\bibitem{land} L. D. Landau, Sov. Phys. JETP {\bf 3}, 920 (1956).

\bibitem{ckz} V. A. Khodel, M. V. Zverev, and J. W. Clark,
\prb {\bf 71}, 012401 (2005).

\bibitem{lanl2} E. M. Lifshitz, L. P. Pitaevskii, {\it Statistical Physics, Part 1},
(Butterworth-Heinemann, Oxford, 1996) \S 58.

\bibitem{sqrt} J. W. Clark, V. A. Khodel, and M. V. Zverev,
JETP Lett. {\bf 81}, 315 (2005).

\bibitem{pr1} V. A. Khodel, V. R. Shaginyan, and
V. V. Khodel, Phys. Rep. {\bf 249}, 1 (1994).

\bibitem{dkss} J. Dukelsky, V. A. Khodel, P. Schuck, and V. R. Shaginyan,
Z. Phys. {\bf 102}, 245 (1997).

\bibitem{ron} R. Lifshitz, Z. Phys. {\bf 51}, 1156 (2011).

\bibitem{prb} V. R. Shaginyan, A. Z. Msezane, K. G. Popov, J. W. Clark,
M. V.  Zverev, and V. A. Khodel, \prb {\bf 86}, 085147 (2012).

\bibitem{pikul} N. Oeschler, S. Hartmann, A. P. Pikul, C. Krellner, C. Geibel, and F.
Steglich, Physica B {\bf 403},  1254 (2008).

\end{thebibliography}
\end{document}